\documentclass[aps,prl,amsmath,amssymb,preprint]{revtex4-2}
\usepackage{graphicx}
\usepackage{amsmath}
\usepackage{amssymb}
\usepackage{color}
\usepackage{float}
\usepackage{soul}

\usepackage{dcolumn}
\usepackage{bm}
\usepackage{xcolor} 
\begin{document}

\section{Title}
All-optical polarization and amplitude modulation of second harmonic generation in atomically thin semiconductors


\section{Author list}
Sebastian Klimmer,$^1$ Omid Ghaebi,$^1$ Ziyang Gan,$^2$ Antony George,$^{2,3}$ Andrey Turchanin,$^{2,3}$ Giulio Cerullo,$^4$ Giancarlo Soavi$^{1,3}$
\section{Affiliations}
$^1$Institute of Solid State Physics, Friedrich Schiller University Jena, 07743 Jena, Germany

$^2$Institute of Physical Chemistry, Friedrich Schiller University Jena, 07743 Jena, Germany

$^3$Abbe Center of Photonics, Friedrich Schiller University Jena, 07745 Jena, Germany

$^4$Dipartimento di Fisica, Politecnico di Milano, 20133 Milan, Italy

\section{Abstract}
\begin{abstract}
Second harmonic generation is of paramount importance in several fields of science and technology, including frequency conversion, self-referencing of frequency combs, nonlinear spectroscopy and pulse characterization. Advanced functionalities are enabled by the modulation of the harmonic generation efficiency, which can be achieved with electrical or all-optical triggers. Electrical control of the harmonic generation efficiency offers large modulation depth at the cost of low switching speed, in contrast to all-optical nonlinear devices which provide high speed and low modulation depth. Here we demonstrate all-optical modulation of second harmonic generation in MoS$_{2}$ with close to 100\,\% modulation depth and speed limited only by the fundamental pulse duration. This result arises from the combination of the D$_{3h}$ crystal symmetry and the deep sub-wavelength thickness of the sample and it can therefore be extended to the whole family of transition metal dichalcogenides, thus providing a large flexibility in the design of advanced nonlinear optical devices such as high-speed integrated frequency converters, broadband autocorrelators for ultra-short pulse characterization and tunable nanoscale holograms.
\end{abstract}

\maketitle
\section{Main text}
Stemming from the first demonstration of optical harmonic generation \cite{franken1961generation}, nonlinear optics has been in the spotlight of science and technology for more than half-century. In particular, second harmonic generation (SHG) is a second-order nonlinear process widely used for frequency conversion, self referencing of frequency combs \cite{hickstein2019self}, crystal symmetry and Rashba effect studies \cite{schmitt2020control,frohna2018inversion}, sensing \cite{tran2017applications}, interface spectroscopy \cite{shen1989optical} and ultra-short pulse characterization \cite{weiner2011ultrafast}. Besides free-space applications, there is an increasing interest towards the realization of micro-scale integrated nonlinear devices. Here, a major challenge comes from the centro-symmetric nature of silicon (Si) and silicon nitride (Si$_3$N$_4$), which forbids second-order nonlinearities. Large efforts have been devoted to the integration of nonlinear crystals such as lithium niobate \cite{wang2017metasurface, chen2019ultra} or to symmetry breaking in Si and Si$_3$N$_4$, for instance via strain \cite{cazzanelli2012second}, electric fields \cite{timurdogan2017electric} or the photogalvanic effect \cite{lu2021efficient}.\\

Two-dimensional (2D) materials, such as graphene and transition metal dichalcogenides (TMDs), hold great promise for nonlinear optical applications. They have strong and broadband optical response \cite{autere2018nonlinear, trovatello2021optical}, combined with the possibility of harmonic generation enhancement at excitonic resonances in TMDs \cite{wang2015giant} and at multi-photon resonances in the graphene's Dirac cone \cite{soavi2018broadband, massicotte2021hot}. In addition, thanks to their flexibility and mechanical strength \cite{ferrari2015science}, they are easy to integrate on photonic platforms. Various functionalised devices for sensing and frequency conversion have been demonstrated on fibers \cite{an2020electrically}, waveguides \cite{alexander2017electrically} and microrings \cite{liu2020low}, while direct patterning of TMDs has been used to realize atomically thin meta-lenses \cite{van2020exciton,lin2020diffraction} and nonlinear holograms \cite{dasgupta2019atomically,lochner2019controlling}. In addition, harmonic generation in 2D materials can be efficiently tuned by external electrical \cite{seyler2015electrical, klein2017electric, soavi2018broadband,soavi2019hot}  or all-optical excitation \cite{taghinejad2020photocarrier,cheng2020ultrafast}, offering an additional degree of freedom for the design of advanced nanoscale devices.\\

However, all the electrical and all-optical schemes that have been proposed to date for SHG modulation in 2D materials have significant downsides. On one hand, electrical modulation has been demonstrated in tungsten diselenide (WSe$_2$) monolayers \cite{seyler2015electrical} by tuning the oscillator strength of neutral and charged exciton resonances through electrostatic doping, and in molybdenum disulfide (MoS$_2$) homo-bilayers \cite{klein2017electric}, by breaking the naturally occurring inversion symmetry through electrical gating, in the latter case with large modulation depth up to a factor of 60. However, electronics is intrinsically slower compared to optics and photonics. On the other hand, all-optical SHG modulation has been achieved by quenching of the exciton oscillator strength following ultrafast optical excitation in MoS$_2$ \cite{taghinejad2020photocarrier,cheng2020ultrafast}. This approach offers high modulation speed, limited in principle only by the excited state/exciton lifetime ($\sim$ tens of ps). However, the largest depth in all-optical SHG modulation reported to date in TMDs is 55\,\% \cite{taghinejad2020photocarrier}, with a strong dependence on the excitation wavelength and fluence. In addition, this scheme for all-optical SHG modulation is only effective for excitation and frequency conversion above-gap or at excitonic resonances, while it is not applicable for below-gap excitation, leading to a naturally limited spectral bandwidth.\\

Here we demonstrate a novel approach for the all-optical control of the second harmonic (SH) polarization in MoS$_2$ and show that this can be used for all-optical modulation of the SH efficiency with modulation depth close to 100\,\% and speed limited only by the fundamental frequency (FF) pulse duration. Our method relies solely on symmetry considerations in combination with the deep sub-wavelength thickness of the sample and thus does not require resonant enhancement or above-gap excitation for its implementation. Moreover, the same approach can be extended to any 2D material belonging to the D$_{3h}$ symmetry group, thus for instance to any material of the TMDs' family. Our findings provide a new strategy for the tuning of the fundamental properties of light (polarization and amplitude) in the nonlinear regime and in the 2D thickness limit, and thus pave the way to the design of novel advanced functionalities in high-speed frequency converters, nonlinear all-optical modulators and transistors \cite{wang1995second,mingaleev2002nonlinear}, interferometric autocorrelators for ultra-short pulse characterization and tunable atomically thin holograms \cite{dasgupta2019atomically}.\\

\subsection*{Nonlinear optical characterization}
\begin{figure*}
\centering
\includegraphics{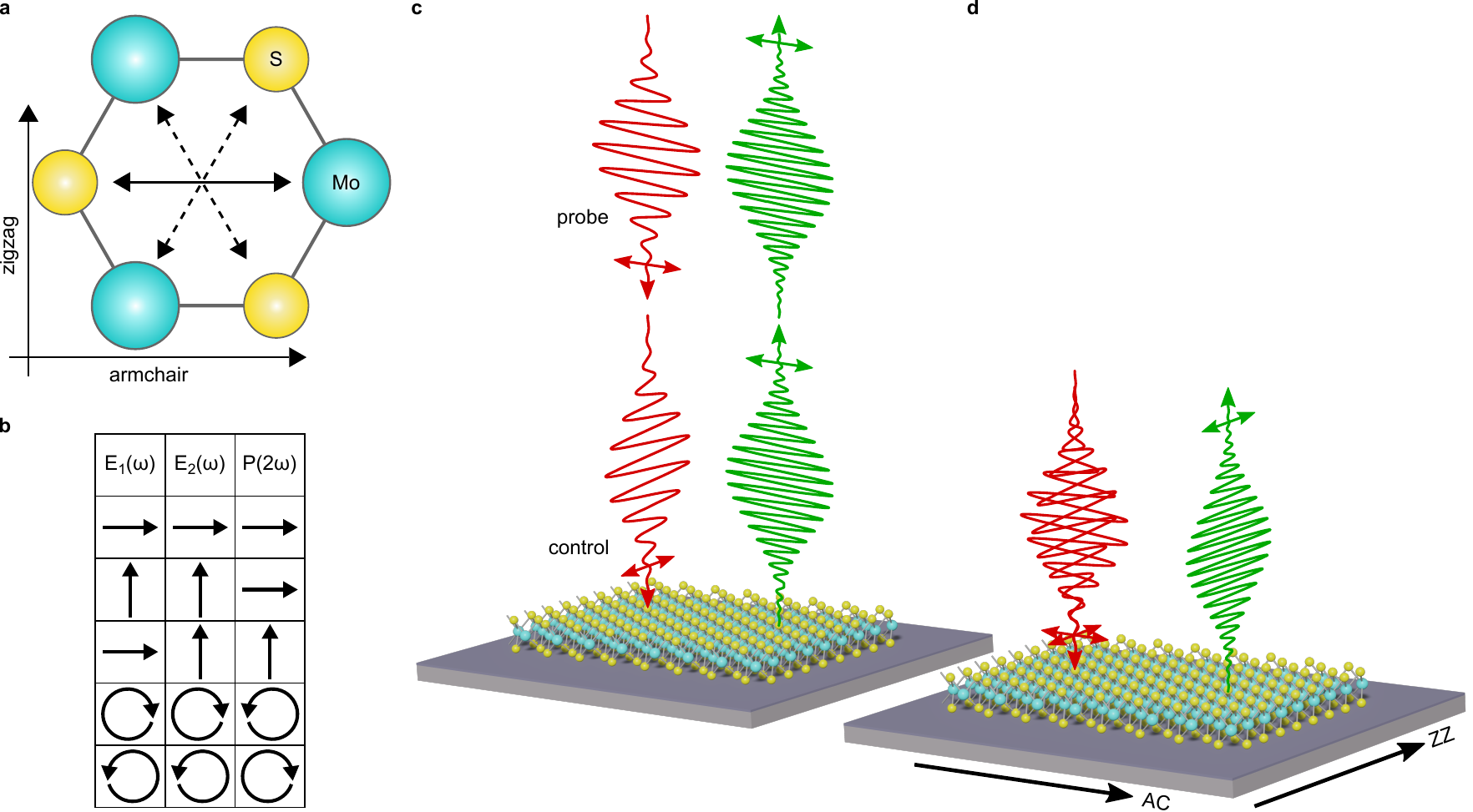}%
\caption{\label{fig:1}\textbf{MoS$_2$ symmetry properties, optical selection rules and all-optical SHG modulation scheme.} (a) Top view of a MoS$_2$ cystal. The arrows inside the hexagon highlight the D$_{3h}$ three-fold rotational symmetry. (b) Schematic of the resulting SH polarization for different combinations of FF along AC (horizontal arrow) and ZZ directions (vertical arrow). (c-d) Sketch of the all-optical SH polarization modulation. The control pulse is polarized along the ZZ direction while the probe pulse is polarized along the AC direction of the MoS$_2$ sample. (c) When the delay between the control and probe pulses is larger than the FF pulse duration, both will generate a SH signal polarized along the AC direction. (d) For zero-delay between the probe and control pulses the SH signal will be emitted along the ZZ direction.}
\end{figure*}

\begin{figure*}
\centering
\includegraphics{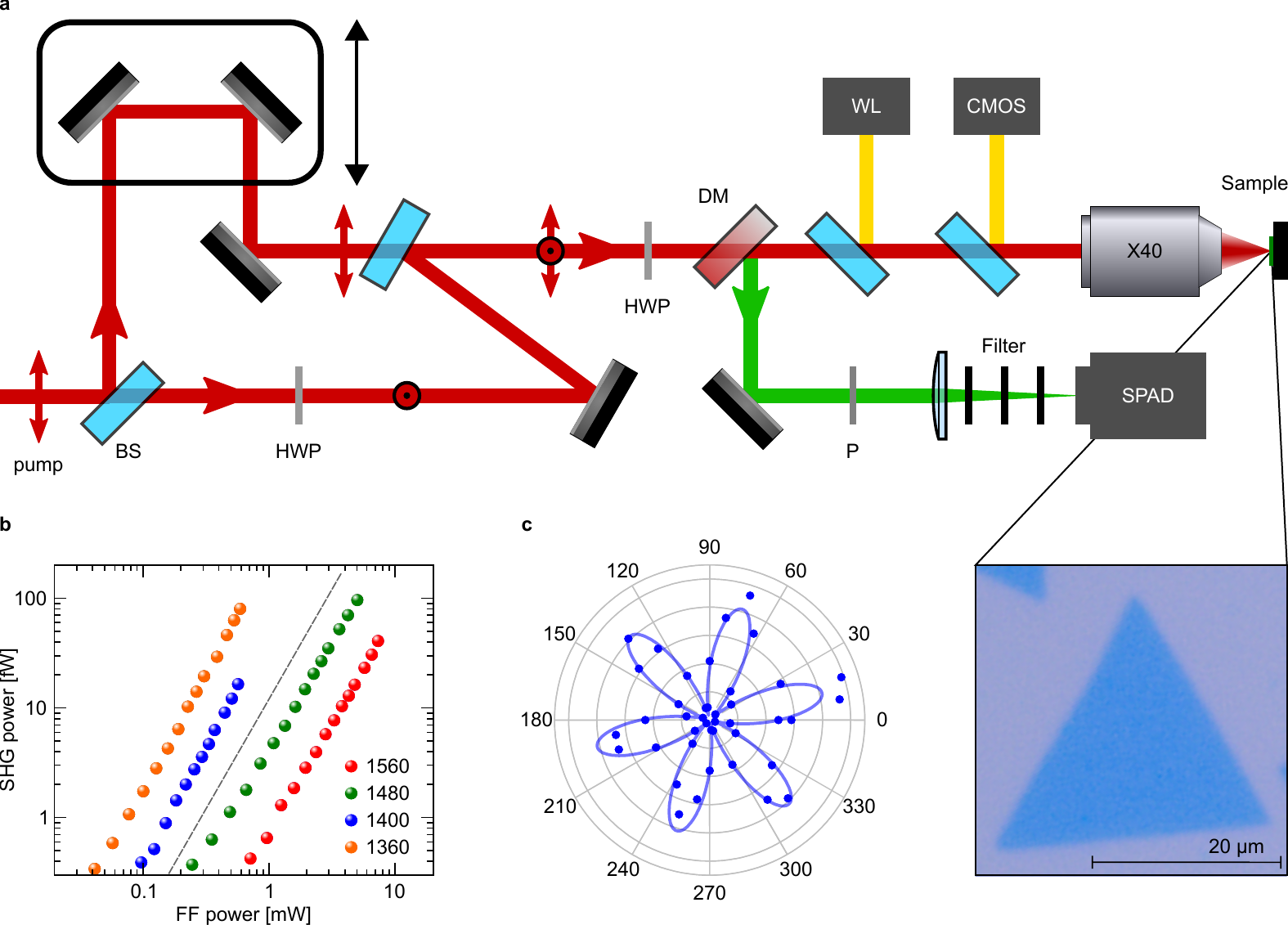}%
\caption{\label{fig:2}\textbf{All-optical modulation set-up and SHG characterization.} (a) Sketch of the setup used for the experiments. For the FF we used an OPO tunable between $\sim$1.3 and 2.0 {\textmu}m. The FF beams are separated into two perpendicular replicas inside a Mach-Zehnder interferometer and subsequently focused onto the MoS$_2$ sample with a 40x microscope objective. The back-scattered SH signal is spectrally filtered and detected with a single-photon avalanche diode. (b) Power dependence of the SHG signal for all wavelengths used in our experiments. The grey dashed line is a guide to the eye to indicate a slope of 2, typical of the SHG process. (c) Polar plot of the normalized SH intensity as a function of the excitation polarization angle $\theta$ with SH polarization detection always parallel to the FF. Blue circles show experimental data and the solid blue line indicates the $\text{cos}^{2}[3(\theta-\phi_{0})]$ fit.}
\end{figure*}

For the experiments, we used high quality monolayer MoS$_2$ flakes, fabricated with a modified chemical vapour deposition method \cite{george2019controlled,shree2019high} on thermally oxidized Si/SiO$_2$ substrates. In contrast to thin films, where surface SHG is allowed due to an out-of-plane component of the second-order susceptibility \cite{boyd2020nonlinear,shen1989optical}, TMDs belong to the D$_{3h}$ symmetry group and thus have only one non-vanishing in-plane component of the nonlinear optical susceptibility (see Methods) \cite{boyd2020nonlinear, malard2013observation,kumar2013second,mennel2018optical,mennel2019second,mennel2020band}.

\begin{eqnarray}
\chi^{(2)} \equiv \chi_{yyy}^{(2)} = -\chi_{yxx}^{(2)} = -\chi_{xyx}^{(2)} = -\chi_{xxy}^{(2)}
\label{eq:one},
\end{eqnarray}

where x and y refer to the in-plane Cartesian coordinates of the SH polarization and of the two FFs. A sketch of the hexagonal lattice for MoS$_2$ is shown in Fig.\;\ref{fig:1}(a), where the Cartesian coordinates are defined with respect to the two main lattice orientations: the armchair (AC) and zigzag (ZZ) directions. In this framework, the SH intensity I$^{2\omega}$ as a function of the FF for any TMD along the AC and ZZ directions can be written as

\begin{eqnarray}
I_{AC}^{2\omega} & \sim &\lvert E_{AC}^{2}-E_{ZZ}^{2} \rvert^{2}
\label{eq:two},
\\
I_{ZZ}^{2\omega} & \sim &\lvert 2E_{AC}E_{ZZ} \rvert^{2}
\label{eq:three},
\end{eqnarray}

where E$_{AC}$ and E$_{ZZ}$ correspond to the FF fields with polarization along the AC and ZZ directions respectively \cite{wen2019nonlinear,woodward2016characterization,rosa2018characterization,mennel2019second}. Thus, the SHG from two electric fields with the same polarization (either along AC or ZZ) will always result in an emitted SH intensity with polarization along the AC direction, as depicted in Fig. \ref{fig:1}(b). This is indeed the case of all the SHG experiments performed to date on 2D materials \cite{wang2015giant,malard2013observation,rosa2018characterization,mennel2019second}. On the other hand, two ultrashort FFs with perpendicular polarization (along the AC and ZZ directions) and with the same amplitude will generate a SH signal along the AC direction if they do not overlap in time (Fig. \ref{fig:1}(c)), while they will generate a SH signal along the ZZ direction at zero-delay (Fig. \ref{fig:1}(d)), thus leading to an ultrafast $90^{\circ}$ polarization switch within the FFs pulse duration. Finally, for circularly polarized FFs, the emitted SH has opposite circular polarization due to valley-dependent selection rules (see analysis of equation (\ref{eq:five})) \cite{zhang2020near, xiao2015nonlinear,saynatjoki2017ultra}.\\

\begin{figure*}
\centering
\includegraphics{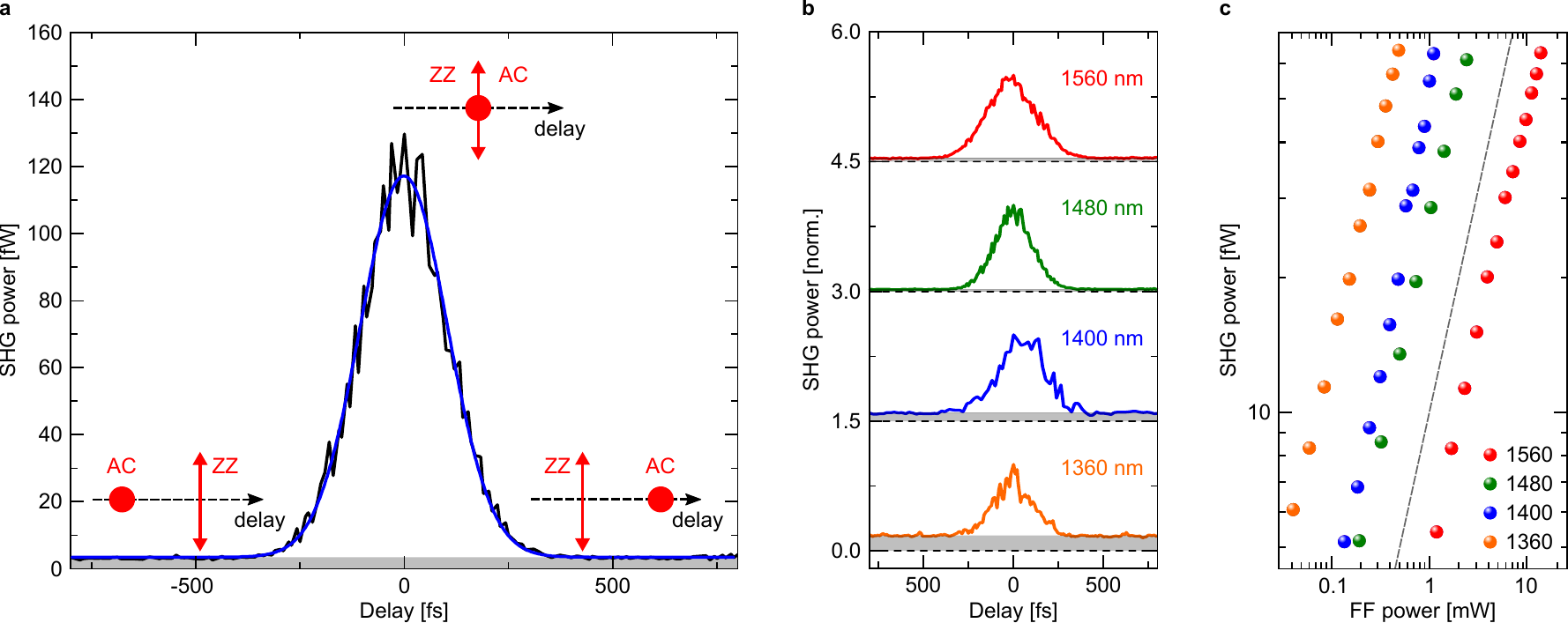}%
\caption{\label{fig:3}\textbf{Ultrafast polarization and amplitude all-optical switching.} (a) SH intensity for 1480\;nm FF wavelength (black line) measured along the zigzag direction as a function of the delay between the two perpendicularly polarized FFs (illustrated in the inset). The blue line corresponds to a Gaussian fit of the autocorrelation curve. The grey shaded area represent the noise level at $\sim$3.5 fW of our experiments. (b) Normalized SH intensity and noise level for all FF wavelengths used in our experiments. (c) Double logarithmic plot of the emitted SH power as a function of the incident FF power along the AC direction. The grey dashed line is a guide to the eye representing a slope of 1.}
\end{figure*}
The SHG measurements were performed with the set-up shown in Fig.\;\ref{fig:2}(a) and described in the Methods. In order to realize all-optical polarization switching and SH modulation it is crucial to first characterize the relative orientation between the FFs and the MoS$_2$ sample. To do so, we first performed SHG experiments using a tunable near-IR optical parametric oscillator (OPO) as a FF. The emitted SHG power for the FF wavelengths used in our experiments (between 1360\;nm and 1560\;nm) is shown in Fig.\;\ref{fig:2}(b). The slope of 2 in the double logarithmic plot is a further proof of a genuine SHG process. The crystal orientation of the MoS$_2$ sample was determined for each FF wavelength by SH polarization dependent experiments with an additional polarizer in front of the detector to measure the SH parallel to the excitation polarization \cite{li2013probing,malard2013observation,rosa2018characterization,mennel2019second}: the SH intensity is proportional to $\text{cos}^{2}[3(\theta-\phi_{0})]$, where $\theta$ is the FF polarization angle and $\phi_{0}$ is the rotation of the AC direction relative to the p-polarization in the laboratory frame. Fig.\;\ref{fig:2}(c) is an example of the SH polar plot for a FF wavelength of 1400\;nm and shows that the AC direction is tilted by $\phi_{0}=13.6^{\circ}+n\cdot60^{\circ}$ ($n$ integer) with respect to p-polarization in the lab coordinates. In addition, Fig.\;\ref{fig:1}(c) confirms the absence of any detectable strain in our sample, since an uniaxial strain would result in a symmetric attenuation along its direction of action \cite{mennel2018optical,mennel2019second}. The small asymmetry of the lobes along the two AC directions ($\sim$\,15$^\circ$/70$^\circ$ and $\sim$\,190$^\circ$/250$^\circ$) is attributed to polarization scrabbling effects due to the use of a dichroic mirror in reflection geometry \cite{mouchliadis2021probing}. Finally, based on the results in Fig.\;\ref{fig:2}(b), we determine the modulus of the complex second-order nonlinear susceptibility at the FF wavelengths used in our experiments. To do so, we estimate the optical losses of the setup from the SH emission at the sample position to the detection on the Single Photon Avalanche Detector (SPAD) and calculate the SH tensor element $\chi^{(2)}$ of MoS$_2$, as described in the Methods. We thus obtained effective second-order susceptibility values for our FF wavelengths 1360\;nm, 1400\;nm, 1480\;nm and 1560\;nm of $\sim$\,282.1\;pm/V, $\sim$\,153.7\;pm/V, $\sim$\,44.7\;pm/V, $\sim$\,24.2\;pm/V respectively. The highest value obtained at 1360\;nm FF wavelength is due to exciton resonant SH enhancement \cite{le2017impact,wang2015giant}. All values are in good agreement with those previously reported by experiments performed at similar FF wavelengths \cite{woodward2016characterization, malard2013observation,clark2015near,le2017impact,daFonseca2021Second} and predicted by theory \cite{trolle2014theory,clark2014strong}. It is worth noting that for single layer TMDs, where interlayer interference \cite{Kim2020second} is absent, SHG is insensitive to the phase of the nonlinear optical response.\\

\begin{figure*}
\centering
\includegraphics{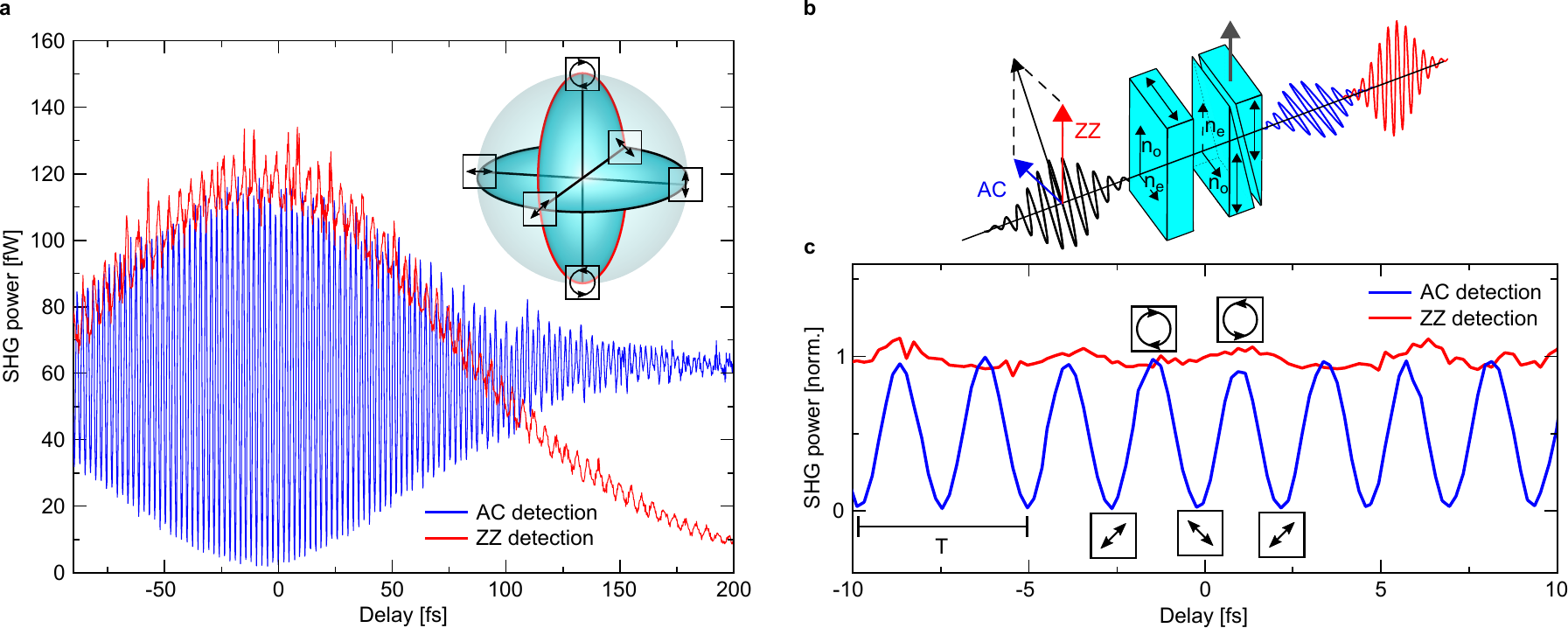}%
\caption{\label{fig:4}\textbf{Phase-locked all-optical SH modulation along armchair and zigzag directions.} (a) SH power for 1480\;nm FF wavelength along AC (blue) and ZZ (red) directions as a function of the relative delay between the two perpendicularly polarized FFs. The inset shows the Poincaré-sphere with polarization directions on its orthodromes. The red line indicates the change in polarization when tuning the delay between the two FFs with a birefringent delay line. (b) Schematic image of the modified common-path delay generator. The two perpendicular components of an incident pulse, polarized at 45$^{\circ}$ with respect to the AC/ZZ directions, are delayed by two birefringent alpha barium borate crystals with perpendicular optical axis, whereby the variable thickness of the second one allows to control the delay with interferometric precision. (c) Zoom-in of (a), showing the detected SH power for 1480\;nm FF along the AC (blue) and ZZ directions (red) close to zero delay.}
\end{figure*}
\subsection*{Nonlinear all-optical modulation}
Having defined the AC and ZZ directions of our sample, we now demonstrate all-optical SH polarization and amplitude modulation. We separate the FF beam into two perpendicular replicas, align them along the sample's AC and ZZ directions using a half-waveplate and control their relative delay with a motorized mechanical stage (see Methods for details). For large delays (\textit{i.e.}, longer than the FF pulse duration) between the two perpendicular FFs, SH will be emitted by each individual FF along the AC direction following equation (\ref{eq:two}). Instead, at zero-delay, when the two FFs overlap perfectly in time, the SH intensity along the AC direction will go to zero and the SH signal will be emitted only along the ZZ direction. Fig.\;\ref{fig:3}(a) shows the SH average power emitted along the ZZ direction as a function of the delay between the two perpendicularly polarized FFs and for a FF wavelength of 1480\;nm. The Gaussian fit, blue curve in Fig.\;\ref{fig:3}(a), has a FWHM of $\sim$\,250\;fs, corresponding to the autocorrelation function of our OPO pulse with duration of $\sim$\,175\;fs. Moreover, the SH signal along the ZZ direction is now ideally background free, demonstrating the potential of ultrafast SH polarization switching and close to 100\,\% amplitude modulation of our approach.\\

We further note that this result is solely based on symmetry considerations and thus provides ultra-broadband response, not limited to above-gap or resonant exciton pumping. Indeed, we obtained the same result for all the FF wavelengths used in our experiments, as shown in Fig.\;\ref{fig:3}(b). The possibility to emit SH along perpendicular directions (AC and ZZ) with the same efficiency is a unique feature that arises from the combination of symmetry and deep sub-wavelength thickness of TMDs, which relaxes the phase-matching constraints typical of harmonic generation. This result could have an immediate application in background free ultra-broadband and ultra-short pulse characterization. For instance, in the most advanced commercial systems \cite{ape} for ultra-short pulse characterization one has to switch between collinear and non-collinear geometries in order to collect either the interferometric or the background-free intensity autocorrelation signals respectively. In contrast, in our approach both signals are accessible using the same geometry and by simply switching the SH detection from AC to ZZ. Further, following equation (\ref{eq:three}), the power scaling of the emitted SH along the ZZ direction is linear with respect to each of the FF intensities. This is confirmed by the power-dependent SHG measurements reported in Fig.\;\ref{fig:3}(c), where we show the emitted SH power along the ZZ direction at zero delay between the two FFs and as a function of the AC polarized FF power.\\

To gain more insight into the temporal evolution of the emitted SH polarization and amplitude, we finally scan the delay between the two perpendicularly polarized FFs with interferometric precision and measure the emitted SH along both AC and ZZ directions, as shown in Fig.\;\ref{fig:4}(a). In order to control the delay between two perpendicular pulses with the desired sub-optical cycle precision we used the common-path delay generator sketched in Fig.\;\ref{fig:4}(b) and described in the Methods. As expected, for delays longer than our pulse duration the SH power is emitted only along the AC direction, blue curve in Fig.\;\ref{fig:4}(a), and no signal is detected along the ZZ direction, red curve in Fig.\;\ref{fig:4}(a). Instead, for delays close to zero we observe a strong ultrafast modulation of the SH power emitted along the AC direction. This can be better appreciated looking at Fig.\;\ref{fig:4}(c), which shows the emitted SH power along AC and ZZ directions at 1480\;nm for delays between -10\;fs and +10\;fs.

It is useful to note that for delays much shorter compared to the pulse duration our interferometric measurement is the analogue of tuning the polarization of one single FF pulse along the orthodrome of the Poincaré-sphere, inset in Fig.\;\ref{fig:4}(a). This corresponds to a rotation of the FF polarization from $-45^{\circ}$ with respect to the AC-ZZ directions (at zero delay) to left/right circular polarization (at delay $\tau = \pm\frac{T}{4}$, where $T$ is the FF optical cycle) to $+45^{\circ}$ with respect to the AC-ZZ directions at delay $\tau = \pm\frac{T}{2}$. This result is consistent with the theoretical SH polarization $\textbf{P}^{(2)}$ generated by an arbitrary elliptically polarized FF \cite{saynatjoki2017ultra} after a simple basis transformation to account for the rotation by $-45^{\circ}$ with respect to AC/ZZ directions:

\begin{eqnarray}
\textbf{P}^{(2)} = \epsilon_{0}\chi^{(2)}\lvert\textbf{E}\rvert^{2}\begin{pmatrix} \widehat{ZZ} \pm i\,\text{sin}(2\vartheta)\widehat{AC}\end{pmatrix}
\label{eq:five},
\end{eqnarray}

Here $\vartheta = 0^{\circ}$ denotes a linearly polarized FF at $45^{\circ}$ with respect to the AC/ZZ direction and $\vartheta = 45^{\circ}$ corresponds to a circularly polarized FF. This clearly shows that the SH component emitted along the AC direction oscillates with a period of $\frac{T}{2}$ as a function of the FF polarization, in contrast to the SH emitted along the ZZ direction. This underpins the interferometric precision required in order to fully capture the modulation along the AC direction. The experimental results show a weak modulation also for the SH emitted along the ZZ direction, although this is not expected from theory. This could arise from weak strain (\textit{i.e.}, below the limit detectable by our SHG polarization measurements \cite{liang2017monitoring,mennel2019second}), small deviations in the alignment of the detection polarizer with respect to the AC/ZZ directions or from additional terms in the $\chi^{(2)}$ arising from the valley degree of freedom \cite{ho2020measuring, mouchliadis2021probing}. Looking at Fig.\;\ref{fig:4}(c) one can indeed appreciate that at zero delay (FF at $-45^{\circ}$ with respect to AC/ZZ directions) the SH is emitted only along the ZZ direction, while at $\frac{T}{4}$ delay the emitted SH components along AC and ZZ are identical, as expected for circular polarization.\\

\subsection*{Discussion}

In conclusion, we have demonstrated all-optical polarization switching and amplitude modulation of SHG in MoS$_2$. Our approach surpasses all previously reported electrical and all-optical attempts of SH tuning in terms of modulation depth and speed, providing $90^{\circ}$ polarization switch, close to 100\,\% modulation depth and speed limited only by the FF pulse duration. Moreover, our method is intrinsically broadband since it only relies on the crystal symmetry of TMDs. We thus foresee a direct impact of our results on a variety of photonic devices, such as high-speed frequency converters, nonlinear all-optical modulators and transistors \cite{wang1995second,mingaleev2002nonlinear}, autocorrelators for ultra-short pulse characterization and atomically thin optically tunable nonlinear holograms \cite{dasgupta2019atomically}.

\section{Methods}

\textbf{Polarization dependent SH intensity.} The vectorial components of the second-order polarization $P^{(2)}(2\omega)$ for a material with D$_{3h}$ symmetry (such as TMDs) are given by
\begin{eqnarray*}
\begin{pmatrix} P^{(2)}_x(2\omega) \\ P^{(2)}_y(2\omega) \\ P^{(2)}_z(2\omega) \end{pmatrix} = 
\epsilon_0 \begin{pmatrix} 0 & 0 & 0 & 0 & 0 & \chi^{(2)}_{xxy} \\ \chi^{(2)}_{yxx} & \chi^{(2)}_{yyy} & 0 & 0 & 0 & 0 \\ 0 & 0 & 0 & 0 & 0 & 0 \end{pmatrix} \begin{pmatrix} E^2_x(\omega) \\ E^2_y(\omega) \\ E^2_z(\omega) \\ 2E_y(\omega) E_z(\omega) \\ 2E_x(\omega) E_z(\omega) \\ 2E_x(\omega) E_y(\omega) \end{pmatrix}
\end{eqnarray*}\\
where $\chi^{(2)}_{yyy} = -\chi^{(2)}_{xxy} = -\chi^{(2)}_{yxx} = \chi^{(2)}$. If we now consider a TMD oriented in such way that the ZZ and AC directions lie along the x and y Cartesian coordinates respectively and we neglect the z (out-of-plane) direction, we obtain the following expression:
\begin{eqnarray*}
\begin{pmatrix} P^{(2)}_{ZZ}(2\omega) \\ P^{(2)}_{AC}(2\omega) \end{pmatrix} = 
\epsilon_0  \chi^{(2)}(2\omega,\omega,\omega) \begin{pmatrix} 2E_{ZZ}(\omega) E_{AC}(\omega) \\ E^2_{ZZ}(\omega) - E^2_{AC}(\omega)\end{pmatrix}
\end{eqnarray*}\\
Finally, since the SH intensity is proportional to the absolute square value of the second-order polarization, we retrieve equations (\ref{eq:two}) and (\ref{eq:three}) of the main text:
\begin{eqnarray*}
I_{AC}^{2\omega} = \lvert P^{(2)}_{AC} \rvert ^2 \sim  \lvert E^2_{AC} - E^2_{ZZ} \rvert ^2\\
I_{ZZ}^{2\omega} = \lvert P^{(2)}_{ZZ} \rvert ^2 \sim  \lvert 2E_{AC}E_{ZZ} \rvert ^2
\end{eqnarray*}\\

\textbf{SHG setup.} For the FF we used an OPO (Levante IR from APE) pumped by an ytterbium-doped mode locked laser (FLINT12, Light Conversion) with a repetition rate of 76 MHz and 100\;fs pulse duration and generating pulses tunable between $\sim$1.3\;{\textmu}m and 2.0\;{\textmu}m. The FF is then separated into two perpendicular replicas whose relative delay is tuned with two different approaches: a computer controlled motorized stage (M-414.2PD, PI) in a Mach-Zehnder interferometer configuration and a commercial common-path birefringent interferometer (GEMINI, NIREOS) \cite{preda2016linear}. Compared to standard home-built interferometers, the GEMINI provides sub-wavelength interferometric stability with precise control on the delay between the two replicas with attosecond precision. The polarization of the FFs was tuned using a half-waveplate (AHWP05M-1600, Thorlabs) and the power on the sample was controlled by two polarizers (LPNIR050 and WP12L-UB, both Thorlabs). Finally, the two collinear and perpendicularly polarized FFs were focused on the sample using a custom built microscope equipped with a 40x reflective objective (LMM-40X-P01, Thorlabs). The back-scattered SH signal is spectrally separated using a dichroic mirror (DMSP950, Thorlabs), further spectrally purified by filters (FELH0650, FESH850, FESH0950, all Thorlabs) and detected with a single-photon avalanche diode (C11202-050, Hamamatsu). The SH polarization was measured using a wire grid polarizer (WP12L-UB, Thorlabs).\\

\textbf{Estimate of the optical losses of the setup.} In order to quantify the SH signal generated directly at the sample position, optical losses of the different components of the setup must be considered. While the transmission coefficients for the investigated SH wavelengths of the filters and the dichroic mirror (all \textgreater\,96\,\%) were taken from the manufacturer's website, the values for polarizers and the microscope objective were determined experimentally. A transmission of $\sim$\,79\,\% was determined for the wire grid polarizer while for the reflective objective we determined a transmission of 50\,\%. Last, the responsivity of the single-photon avalanche diode was taken into account, which ranges depending on the investigated SH wavelength between $\sim$\,17\,\% and $\sim$\,31\,\%. In total, we estimated our optical losses from the SH emission to the detector to be $\sim$\,86-92\,\% depending on the wavelength.\\

\textbf{Calculation of the second-order nonlinear susceptibility.} The sheet-SH tensor element $\chi_S^{(2)}$ can be calculated from the FF and SH average powers using the equation \cite{woodward2016characterization}:
\begin{eqnarray}
\chi_{S}^{(2)}=\sqrt{\frac{c^{3}\epsilon_{0}f\pi r^{2}t_{FWHM}(1+n_{2})^{6}P_{SHG}(2\omega)}{16 \sqrt{2} S \omega^{2} P_{FF}^{2} (\omega)}}
\label{eq:four},
\end{eqnarray}
where $c$ is the speed of light, $\epsilon_{0}$ is the permittivity of free-space, $f$ = 76 MHz is the pump laser repetition rate, $r \sim$\,1.85\;{\textmu}m  is the focal spot radius, $t_{FWHM} \sim$\,200\;fs is the pulse full width at half maximum, $n_{2} \sim$\,1.45 is the substrate refractive index, $S=0.94$ is a shape factor for Gaussian pulses, $\omega$ is the FF angular frequency, and $P_{SHG}(2\omega)$ - $P_{FF} (\omega)$ are the SH and FF average powers. The effective bulk-like second-order susceptibility $\chi_{eff}^{(2)}$ can be subsequently calculated from equation (\ref{eq:four}) as $\chi_{eff}^{(2)} = \frac{\chi_{S}^{(2)}}{d_{MoS_{2}}}$, where the thickness $d_{MoS_{2}}$ of MoS$_2$ is 0.65\;nm \cite{ferrari2015science,saynatjoki2017ultra}.\\

\textbf{Pulse duration of the FFs.} In order to prove that our method is solely limited by the pulse duration of the FFs, we performed a standard characterization of temporal profile of our OPO source at different wavelengths (Extended Data Fig.\;1). For the measurements we used a home built autocorrelator based on a Michelson interferometer equipped with a motorized and computer controlled translation stage (HPS60-20X-M5, Optosigma). Two identical and temporally delayed replicas of the OPO pulse were then focused onto a 1\;mm thick $\beta$-barium borate crystal (BBO-652H, Eksma Optics) in non-collinear geometry and the background free SHG autocorrelation intensity was detected on a Si photodetector (DET10A2, Thorlabs). From this we obtained values for the autocorrelation in the range of 217\;fs to 310\;fs with gaussian fits, corresponding to pulse durations between 150\;fs and 220\;fs.

\section{Acknowledgments}
The authors greatly acknowledge Habib Rostami and Fabrizio Preda for helpful discussions. This work was supported by the European Union’s Horizon 2020 Research and Innovation program under Grant Agreement GrapheneCore3 881603 (G.S.). This publication is part of the METAFAST project that received funding from the European Union's Horizon 2020 Research and Innovation program under Grant Agreement No. 899673 (G.S. and G.C.). The authors acknowledge the German Research Foundation DFG (CRC 1375 NOA project numbers B2 (A.T.) and B5 (G.S.)) and the Daimler und Benz foundation for financial support (G.S.).

\section{Author Contributions Statement}
 S.K. and G.S. conceived the experiments. S.K. and O.G. performed the all-optical modulation measurements. Z.G., A.G. and A.T. fabricated and provided the high quality MoS$_2$ sample. S.K., G.C and G.S. wrote the manuscript with contribution from all the authors. All the authors participated to the discussion and commented on the manuscript.

\section{Competing Interests Statement}
The authors declare no competing interest.

\section{Figure Legends/Captions (for main text figures)}

\textbf{MoS$_2$ symmetry properties, optical selection rules and all-optical SHG modulation scheme.} (a) Top view of a MoS$_2$ cystal. The arrows inside the hexagon highlight the D$_{3h}$ three-fold rotational symmetry. (b) Schematic of the resulting SH polarization for different combinations of FF along AC (horizontal arrow) and ZZ directions (vertical arrow). (c-d) Sketch of the all-optical SH polarization modulation. The control pulse is polarized along the ZZ direction while the probe pulse is polarized along the AC direction of the MoS$_2$ sample. (c) When the delay between the control and probe pulses is larger than the FF pulse duration, both will generate a SH signal polarized along the AC direction. (d) For zero-delay between the probe and control pulses the SH signal will be emitted along the ZZ direction.

\textbf{All-optical modulation set-up and SHG characterization.} (a) Sketch of the setup used for the experiments. For the FF we used an OPO tunable between $\sim$1.3 and 2.0 {\textmu}m. The FF beams are separated into two perpendicular replicas inside a Mach-Zehnder interferometer and subsequently focused onto the MoS$_2$ sample with a 40x microscope objective. The back-scattered SH signal is spectrally filtered and detected with a single-photon avalanche diode. (b) Power dependence of the SHG signal for all wavelengths used in our experiments. The grey dashed line is a guide to the eye to indicate a slope of 2, typical of the SHG process. (c) Polar plot of the normalized SH intensity as a function of the excitation polarization angle $\theta$ with SH polarization detection always parallel to the FF. Blue circles show experimental data and the solid blue line indicates the $\text{cos}^{2}[3(\theta-\phi_{0})]$ fit.

\textbf{Ultrafast polarization and amplitude all-optical switching.} (a) SH intensity for 1480\;nm FF wavelength (black line) measured along the zigzag direction as a function of the delay between the two perpendicularly polarized FFs (illustrated in the inset). The blue line corresponds to a Gaussian fit of the autocorrelation curve. The grey shaded area represent the noise level at $\sim$3.5 fW of our experiments. (b) Normalized SH intensity and noise level for all FF wavelengths used in our experiments. (c) Double logarithmic plot of the emitted SH power as a function of the incident FF power along the AC direction. The grey dashed line is a guide to the eye representing a slope of 1.

\textbf{Phase-locked all-optical SH modulation along armchair and zigzag directions.} (a) SH power for 1480\;nm FF wavelength along AC (blue) and ZZ (red) directions as a function of the relative delay between the two perpendicularly polarized FFs. The inset shows the Poincaré-sphere with polarization directions on its orthodromes. The red line indicates the change in polarization when tuning the delay between the two FFs with a birefringent delay line. (b) Schematic image of the modified common-path delay generator. The two perpendicular components of an incident pulse, polarized at 45$^{\circ}$ with respect to the AC/ZZ directions, are delayed by two birefringent alpha barium borate crystals with perpendicular optical axis, whereby the variable thickness of the second one allows to control the delay with interferometric precision. (c) Zoom-in of (a), showing the detected SH power for 1480\;nm FF along the AC (blue) and ZZ directions (red) close to zero delay.

\section{Data Availability}
The data that support the plots within this paper and other findings of this study are available from the corresponding author on reasonable request. Source data are provided with this paper.

\end{document}